\documentclass[conference]{IEEEtran}
\IEEEoverridecommandlockouts
\usepackage{cite}
\usepackage{amsmath,amssymb,amsfonts}
\usepackage{mathrsfs}
\usepackage{graphicx}
\usepackage{textcomp}
\usepackage{xcolor}
\usepackage{subfigure}
\usepackage[vlined,boxed,ruled,linesnumbered]{algorithm2e}
\SetKwProg{Fn}{Function}{}{}
\usepackage{algpseudocode}
\def\BibTeX{{\rm B\kern-.05em{\sc i\kern-.025em b}\kern-.08em
    T\kern-.1667em\lower.7ex\hbox{E}\kern-.125emX}}

\newtheorem{problem}{Problem}

\newtheorem{assumption}{Assumption}
\newtheorem{definition}{Definition}

\newtheorem{remark}{Remark}

\begin{document}

\title{A Data-Driven Hybrid Automaton Framework to Modeling Complex Dynamical Systems\\
\thanks{This research was in part supported by the National Science Foundation, under NSF CAREER Award 2143351, and NSF CNS Award no. 2223035.}
}

\author{\IEEEauthorblockN{Yejiang Yang}
\IEEEauthorblockA{\textit{School of Electrical Engineering} \\
\textit{Southwest Jiaotong University}\\
Chengdu, China \\
yangyejiang@my.swjtu.edu.cn}
\and
\IEEEauthorblockN{Zihao Mo}
\IEEEauthorblockA{\textit{School of Computer and Cyber Sciences} \\
\textit{Augusta University}\\
Augusta, USA \\
zmo@augusta.edu}
\and
\IEEEauthorblockN{Weiming Xiang}
\IEEEauthorblockA{\textit{School of Computer and Cyber Sciences} \\
\textit{Augusta University}\\
Augusta, USA \\
wxiang@augusta.edu}
}

\maketitle

\begin{abstract}
 In this paper, a computationally efficient data-driven hybrid automaton model is proposed to capture unknown complex dynamical system behaviors using multiple neural networks. The sampled data of the system is divided by valid partitions into groups corresponding to their topologies and based on which, transition guards are defined. Then, a collection of small-scale neural networks that are computationally efficient are trained as the local dynamical description for their corresponding topologies. After modeling the system with a neural-network-based hybrid automaton, the set-valued reachability analysis with low computation cost is provided based on interval analysis and a split and combined process. At last, a numerical example of the limit cycle is presented to illustrate that the developed models can significantly reduce the computational cost in reachable set computation without sacrificing any modeling precision. 
\end{abstract}


\section{Introduction}

Data-driven methods such as neural networks are widely used in modeling for their effectiveness without relying on the explicit mathematical model or prior knowledge of the system in a variety of research activities, e.g., modeling nonlinear dynamical systems in the description of Ordinary Differential Equations (ODEs) \cite{kuptsov2021artificial} such as modeling thermal conductivity of water-based nanofluid containing magnetic copper nanoparticles in \cite{ghazvini2020experimental}, modeling groundwater-level variation in coastal aquifers in \cite{roshni2020neural}, etc. However, due to the high complexity of large-scale neural network models, some computationally expensive tasks such as reachability analysis are difficult to perform on neural-network-based models. Therefore, computationally efficient modeling methods are in critical need for neural-network-based models.

To improve the performance of the model and ensure that the model is matching the characteristics of the system, e.g., robustness, stability, etc, modeling is becoming a challenging task. The training of neural network models has received particular attention in the machine learning community. For instance, adding Lyapunov constraints in the training neural networks to enhance stabilization of learning nonlinear system in \cite{neumann2013neural}, studying the adversarial robustness of neural networks using robust optimization in \cite{he2020learnability}, utilizing the idea of robust optimization in the training of neural networks in \cite{yang2021robust}, estimating the Lipschitz constant of neural networks in \cite{fazlyab2019efficient}, etc. Besides training, the verification for neural networks plays a crucial part in the usability and safety of the neural-network-based model is investigated in research such as providing reachable set estimation and safety verification for multi-layer neural networks in \cite{xiang2018output}, verifying the neural network model by the star set based set-valued reachability analysis in \cite{tran2020nnv}, providing bound propagation-based neural network verifiers in \cite{zhang2018efficient,wang2021beta}, etc. Training and verification are related to the size of the neural network, i.e., a single neural network that aims to be trained with all samples may lead to the training and verification for the neural network model becoming complex and time-consuming works. 

Inspired by \cite{beg2017model}, a dynamical system can be modeled by a hybrid automaton with a finite number of local topologies plus transitions among them. Furthermore, if the dynamical description for each topology of the hybrid automaton is a neural network, and the neural network only needs to approximate the local system dynamics within that topology, which means the size of each neural network can be scaled down compared with using a large-scale neural network approximating the entire system dynamics. As a result, the computational complexity in either training or verification will be reduced and moreover, due to parallel training, the scalability can be further increased. In this paper, a neural-network-based hybrid automaton is proposed to reduce the computation cost in training and verification for modeling the dynamical systems. First, the given region is divided into valid partitions representing topologies of the proposed model, based on which the guards are defined. Sample data are selected into different groups with which the neural networks are trained respectively as the dynamical description of their corresponding topologies. Then, the Mean Square Error (MSE) and analysis of the set-valued reachability of our proposed method are provided. Lastly, a numerical example is given to illustrate the effectiveness of our approach.

The main contributions of this paper lie in the way to model the dynamical system using our computationally efficient neural-network-based hybrid automaton and its set-valued analysis. The neural-network-based hybrid automaton models the system with multiple neural networks with each trained with the sample group corresponding to its topology, which reduces the computational complexity while increasing scalability. The set-valued analysis is given based on interval analysis and a \emph{Split and} \emph{Combined} process.  

The paper is organized as follows: preliminaries and problem formulation are given in Section II. The main result, modeling with neural-network-based hybrid automaton, and the set-valued analysis are given in Section III. In Section IV, a limit cycle modeling example is provided to evaluate our method. Conclusions are given in Section V.  

\section{Preliminaries and Problem Formulation}

The dynamical systems in the paper are in the general discrete-time form of
\begin{align} \label{eq:system}
    x(k+1) = f(x(k),u(k))
\end{align}
where  $x(k) \in \mathbb{R}^{n_x}$ is the system state, and $u(k) \in  \mathbb{R}^{n_u}$
is the system input, respectively. The evolution of system state is governed by $f:\mathbb{R}^{n_x+n_u} \to \mathbb{R}^{n_x}$, which is an unknown nonlinear discrete-time process. In this paper, we aim to develop a novel data-driven, i.e., neural-network-based, modeling method to approximate this unknown nonlinear mapping $f$. 

\begin{assumption} \label{assumption_1}
It is assumed that state $x(k)$ and input
$u(k)$ are all measurable from unknown
nonlinear system (\ref{eq:system}).
\end{assumption}

Under Assumption \ref{assumption_1}, the training data for modeling is defined as follows.

\begin{definition}\label{def_sampled data}
Sampled data $\mathcal{W}=\{w_1,w_2,\cdots,w_L\}$ of an unknown dynamical system (\ref{eq:system}) is a collection of sampled $L$ traces obtained by measurement, where for each trace $w_i$, $i = 1,\ldots,L$, is a finite sequence of time steps and data $(k_{0,i},d_{0,i}),(k_{1,i},d_{1,i}),\cdots,(k_{M_i,i},d_{M_i,i})$ in which
\begin{itemize}
    \item $k_{\ell,i}\in(0,\infty)$ and $k_{\ell+1,i}=k_{\ell,i}+1$, $\forall \ell = 0,1,\ldots,M_i,\ \forall i = 1,2,\ldots,L$.
    \item $d_{\ell,i}=[x^\top_{i}(k_{\ell,i}),~u^\top_{i}(k_{\ell,i})]^\top \in \mathbb{R}^{n_x+n_u}$,
    $\forall \ell = 0,1,\ldots,M_i$, $\forall i = 1,2,\ldots,L$, where $x_{i}(k_{\ell,i}),u_{i}(k_{\ell,i})$ denote the state and input of the system at $\ell$th step for $i$th trace, respectively.
\end{itemize}
\end{definition}

In this paper, it is assumed that there exist sufficient measurable input and state traces available in $\mathcal{W}$ for the data-driven modeling of nonlinear system (\ref{eq:system}). Instead of using one single large-scale neural network, e.g., Deep Neural Networks (DNN), as in most of the existing results \cite{xiang2018output,xiang2020reachable,tran2020nnv} which commonly suffer overly expensive computation in successive use especially for safety verification before deployment on safety-critical systems, we aim to develop a novel neural-network-based hybrid automaton consisting of a family of small-scale neural networks, i.e., shallow neural networks, along with inferred transitions to not only accurately approximate the unknown nonlinear system model $f$, but also hold a great promise in performing computational-efficient verification.

In this work, we consider feedforward neural networks in the form of  $\Phi: \mathbb{R}^{n_0} \to \mathbb{R}^{n_{L}}$ defined by the following recursive equations in the form of 
\begin{align}\label{eq:nn}
\begin{cases}
    \eta_{\ell} = \phi_{\ell}(W_{\ell} \eta_{\ell-1}+b_\ell),~\ell = 1,\ldots,L
    \\
    \eta_{L} = \Phi(\eta_0)
    \end{cases}
\end{align}
where $\eta_\ell$ denotes the output of the $\ell$-th layer of the neural network, and in particular $\eta_0\in\mathbb{R}^{n_0}$ is the input to the neural network and $\eta_L\in \mathbb{R}^{n_L}$ is the output produced by the neural network, respectively. $W_\ell \in \mathbb{R}^{n_{\ell}\times n_{\ell-1}}$ and $b_{\ell} \in \mathbb{R}^{n_{\ell}}$ are weight matrices and bias vectors for the $\ell$-th layer.  $\phi_\ell = [\psi_{\ell},\cdots,\psi_{\ell}]$ is the concatenation of activation functions of the $\ell$-th layer in which $\psi_{\ell}:\mathbb{R} \to \mathbb{R}$ is the activation function. 

In this paper, we focus on reducing the computational complexity of neural-network-based models in terms of reachability analysis. The reachable set of neural network (\ref{eq:nn}) is given as follows.

\begin{definition}\label{def_reachable}
Given neural network (\ref{eq:nn}) with a bounded input set $\mathcal{V}$, the following set
\begin{equation}
\mathcal{Y} \triangleq \{\eta_{L} \mid \eta_{L} = \Phi(\eta_{0}),~ \eta_{0} \in \mathcal{V} \subset \mathbb{R}^{n}\}
\end{equation}
is called the reachable set of neural network (\ref{eq:nn}). 
\end{definition}

\begin{remark}
The computation complexity for reachable set computation of  $\mathcal{Y}$ heavily relies on the number of layers and neurons. To enable computationally efficient neural-network-based models in particular for reachability analysis and successive verification procedures, we have to choose neural networks with fewer layers and neurons, however, it usually leads to low training accuracy for complex nonlinear systems. Our goal in this paper is to overcome this dilemma of computation complexity versus training accuracy. 
\end{remark}

\section{Main Results}

\subsection{Neural-Network-Based Hybrid Automata}

The main goal of this paper is to develop an efficiently verifiable data-driven model for nonlinear system (\ref{eq:system}). Inspired by the hybridization methods for the analysis of nonlinear systems in \cite{asarin2007hybridization,bak2016scalable} which enable the efficient verification procedures of hybrid automata with complex, nonlinear dynamics through an abstraction process of multiple local subsystems of simplified forms, we propose a hybridization method to model unknown nonlinear dynamics (\ref{eq:system}) in data-driven modeling scenarios utilizing a collection of small/moderate-size neural networks characterizing local system behaviors.

A neural-network-based hybrid automaton consists of variable components describing both dynamics and the discrete transition logic of a system, which will be used as the modeling framework in the rest of this paper. 
\begin{definition}\label{nn_hybrid}
A  neural-network-based hybrid automaton is defined by a tuple $\mathcal{H} \triangleq \left \langle \mathcal{Q},\mathcal{X},init,\mathcal{U},\mathcal{E},g,\mathcal{G},inv,\mathcal{F} \right \rangle$ in which the components are defined by:
\begin{itemize}
\item \emph{Topologies}: $\mathcal{Q} \triangleq \{q_1,q_2,\ldots,q_N\}$ is a finite set of \emph{topologies}.

\item \emph{State Variables}: $\mathcal{X} \subset \mathbb{R}^{n_x}$ is the set of state variables with the \emph{state} defined by $(q,x)\in \mathcal{Q}\times \mathcal{X}$.

\item \emph{Initial conditions}: $\emph{init}\subseteq \mathcal{Q}_0\times \mathcal{X}_{0}$ is the initial state set, in which $\mathcal{Q}_0\subseteq \mathcal{Q}$ and $\mathcal{X}_{0}\subseteq \mathcal{X}$.

\item \emph{Inputs}: $\mathcal{U} \triangleq \{u_1,u_2,\ldots,u_M\}$ is the set of inputs for each topologies.

\item \emph{Transitions}: $\mathcal{E}\subset \mathcal{Q} \times \mathcal{Q}$ is the set of transitions where a discrete
transition from $i$th topology to $j$th topology is taking place, i.e.,  $q_i \to q_j$, $e_{ij}=(q_i,q_j)\in \mathcal{E}$.

\item \emph{Guard functions}: $g:\mathcal{E} \to \mathcal{G}$ is the guard function mapping each transition element $e_{ij}$ to its guard $g(e_{ij})\in \mathcal{G}$. 

\item \emph{Guards}: $\mathcal{G}\subseteq2^{\mathcal{X}}$ is the guard set which satisfies $\forall e_{ij}\in \mathcal{E}$, $g(e_{ij})\in \mathcal{G}$. The \emph{guard} is satisfied by the state when the hybrid automaton model takes a transition from the current topology to another given topology, i.e., $(q_k,x_k)\vDash g(e_{ij})$ if and only if $q_k=q_i$ and $x_k\in g(e_{ij})$.

\item \emph{Invariants}: $inv: \mathcal{Q}\to 2^{\mathcal{X}}$ is a mapping that assigns an invariant $inv(q)\subseteq \mathcal{X}$ for each topology $q\in{\mathcal{Q}}$. An \emph{invariant} is satisfied by all the states of a hybrid automata model for a given topology, i.e., $(q,x)\vDash inv(q)$ if and only if $x\in inv(q)$.

\item \emph{Set of Dynamical Description}: $\mathcal{F}$ is the set of dynamical description which describes the dynamical process for each given topology $q\in \mathcal{Q}$. In this paper, neural network $\Phi_q(x(k),u(k)) \in \mathcal{F}$ defines the dynamics for each topology $q\in \mathcal{Q}$ in a given time step $k\in[k_1,k_2]$.
\end{itemize}
\end{definition}

Given sampled data set $\mathcal{W}$ defined by Definition \ref{def_sampled data}, the data-driven modeling problem for unknown dynamical system (\ref{eq:system}) is to establish a neural-network-based hybrid automaton, i.e., $\mathcal{H}=\left \langle \mathcal{Q},\mathcal{X},init,\mathcal{U},\mathcal{E},g,\mathcal{G},inv,\mathcal{F} \right \rangle$ with $\mathcal{F}$ represented by a collection of neural networks $\Phi_q$, $q \in \mathcal{Q}$. In addition, the constructed neural-network-based hybrid automaton model $\mathcal{H}$ is expected to be with low computational complexity without sacrificing training accuracy. Specifically, the modeling and verification challenges are described as follows.
\begin{problem} 
Given sampled data set $\mathcal{W}$ collected from measurable input and state traces of dynamical system (\ref{eq:system}), how does one construct a hybrid automaton $\mathcal{H}$ embedded with neural network in the form of (\ref{eq:nn}) to accurately capture the system dynamical behaviors of system (\ref{eq:system})?
\end{problem}
\begin{problem}
Given neural-network-based hybrid automaton model $\mathcal{H}$  as defined by Definition \ref{nn_hybrid}, how does one perform efficient verification procedures on $\mathcal{H}$?
\end{problem}


\subsection{Data-Driven Modeling Processes}

To learn local system behaviors, state space partitioning is required for the segmentation of training data set $\mathcal{W}$. 
\begin{definition}\label{def_division}
Given a compact set $\mathcal{X}\subset\mathbb{R}^{n_x}$. A finite collection of sets $\mathscr{X}\triangleq\{\mathcal{X}^{(1)},\mathcal{X}^{(2)},\ldots,\mathcal{X}^{(N)}\}$ is called a partition of $\mathcal{X}$ if (1) $\mathcal{X}^{(q)} \subseteq \mathcal{X}$, $\forall q = 1,\ldots,N$; (2) $\mathrm{int}(\mathcal{X}^{(q)})\cap \mathrm{int}(\mathcal{X}^{(p)})=\emptyset, \forall q \ne p$; (3) $\mathcal{X}\subseteq \bigcup_{q=1}^{N} \mathcal{X}^{(i)}$. Each elements of $\mathcal{X}^{(q)}$ of partition $\mathscr{X}$ is called a cell.    
\end{definition}

\begin{remark}
In this paper, we use cells defined by hyper-rectangles which are given as follows: For any bounded state set $\mathcal{X} \subseteq \mathbb{R}^{n_x}$, we define $\mathcal{X} \subseteq \bar{\mathcal{X}}$, where $\bar{\mathcal{X}} = \{x \in \mathbb{R}^{n_x} \mid \underline{x} \le x \le \bar{x}\}$, in which $\underline{x}$ and $\bar{x}$ are defined as the lower and upper bounds of state $x$ in $\mathcal{X}$ as  $\underline{x}=[\mathrm{inf}_{x\in \mathcal{X}}(x_1),\ldots,\mathrm{inf}_{x\in \mathcal{X}}(x_{n_x})]^{\top}$ and $\bar{x}=[\mathrm{sup}_{x\in \mathcal{X}}(x_1),\ldots,\mathrm{sup}_{x\in \mathcal{X}}(x_{n_x})]^{\top}$, respectively. Then, we are able to partition interval $\mathcal{I}_i=[\mathrm{inf}_{x\in \mathcal{X}}(x_i),~\mathrm{sup}_{x\in \mathcal{X}}(x_i)]$, $i \in \{1,\ldots,n_x\}$ into $N_i$ segments as $\mathcal{I}_{i,1}=[x_{i,0},x_{i,1}]$, $\mathcal{I}_{i,2}=[x_{i,1},x_{i,2}]$, $\ldots$, $\mathcal{I}_{i,N_i}=[x_{i,N_i-1},x_{i,N_i}]$, where $x_{i,0}=\mathrm{inf}_{x\in \mathcal{X}}(x_i)$, $x_{i,N_i}=\mathrm{sup}_{x\in \mathcal{X}}(x_i)$ and $x_{i,n} = x_{i,0} + \frac{m(x_{i,N_i}-x_{i,0})}{N_i}$, $m \in \{0,1,\ldots,N_i\}$. The cells then can be constructed as $\mathcal{X}_q = \mathcal{I}_{1,m_1} \times\cdots\times \mathcal{I}_{n_x,m_{n_x}}$, $q \in \{1,2,\ldots,\prod\nolimits_{s=1}^{n_x}N_s\}$, $\{m_1,\ldots,m_{n_x}\}\in \{1,\ldots,N_1\} \times \cdots \times \{1,\ldots,N_{n_x}\}$. To remove redundant cells, we have to check if the cell has an empty intersection with $\mathcal{X}$. Cell $\mathcal{X}^{(q)}$ should be removed if $\mathcal{X}_q \cap \mathcal{X} = \emptyset$, and the remaining cells $\mathcal{X}^{(q)}$ constitute $\mathscr{X} = \{\mathcal{X}^{(1)},\mathcal{X}^{(2)},\ldots,\mathcal{X}^{(N)}\}$.
\end{remark}

Based on a collect of sets $\mathscr{X} = \{\mathcal{X}^{(1)},\mathcal{X}^{(2)},\ldots,\mathcal{X}^{(N)}\}$, we will be able to segment sampled data set $\mathcal{W}$ into a collection of sets $\mathscr{W} = \{\mathcal{W}_1,\mathcal{W}_2,\ldots,\mathcal{W}_N\}$ where
\begin{align}
    \mathcal{W}_q \triangleq \{x(k) \mid x(k)\in \mathcal{X}_q, ~x(k+1)\in \mathcal{X}^{(q)}\}
\end{align}
in which $x(k)$, $x(k+1)$ are any sampled states in traces $w_i$, $i=1,\ldots,L$ defined by Definition \ref{def_sampled data}. Therefore, $\mathcal{W}_q$ contains all the sampled state traces evolving within cell $\mathcal{X}^{(q)}$. 

To train neural network $\Phi_q$ to model local system behaviors of dynamical system (\ref{eq:system}) evolving in $\mathcal{X}^{(q)}$, the training input-output pairs need to be abstracted from trace set $\mathcal{W}_q$.

\begin{definition}
Given sampled trace $w_i$ in set $\mathcal{W}_q$, an input-output pair is defined as
\begin{align}
    p_{\ell,i,q} = \{d_{\ell,i},~ x_i(k_{\ell+1},i)\}
\end{align}
where $d_{\ell,i}=[x^\top_{i}(k_{\ell,i}),~u^\top_{i}(k_{\ell,i})]^\top$ is given in Definition \ref{def_sampled data}, and $d_{\ell,i} \in \mathcal{W}_q \times \mathcal{U}$, $x_i(k_{\ell+1},i) \in \mathcal{X}_q$. The training data set out of $\mathcal{W}_q$ includes all the input-output pairs and is given as below: 
\begin{align}
    \mathcal{P}_q = \{p_{0,1,q},\ldots,p_{\ell,i,q},\ldots,p_{M_L,L,q}\}
\end{align}
where $\ell = 0,1,\ldots,M_i$, $i = 1,\ldots,L$.  
\end{definition}

Under Assumption \ref{assumption_1}, we assume that there exist a sufficient number of training input-output pairs in each $\mathcal{P}_q$ out of segmented data set $\mathcal{W}_q$ to train neural network $\Phi_q$ for modeling local system behaviors in cell $\mathcal{X}^{(q)}$. 

With training set $\mathcal{P}_q$ for each cell $\mathcal{X}^{(q)}$, the neural networks can be trained respectively for location $q$. The set of dynamical description $\mathcal{F}$ which is a collection of neural networks $\Phi_q$, $q = 1,\ldots,N$ can be trained, which can be summarized as the following problem: 
\begin{align}\label{eq:opt}
    \min\nolimits_{W_q,b_q}\left\|\Phi_q(D_q)-\hat{X}_q\right\|,~q = 1,2,\cdots,N
\end{align}
where $W_q$ and $b_q$ are weight matrices and bias vectors to determine neural network $\Phi_q$, $D_q$ are input data matrix and $\hat{X}_q$ is output data matrix from input-output pair $\mathcal{P}_q$, respectively. 

\begin{remark}
In general, the learning processes can be viewed as an optimization procedure to find optimized weights and biases that minimize the error between the output of the trained neural network and output training data, as described in (\ref{eq:opt}).  Instead of using one single neural network for modeling, a collection of neural networks $\Phi_q$, $q=1,\ldots, N$ are used to model local system behaviors in each cell $\mathcal{X}^{(q)}$, $q=1,\ldots, N$. We have the following advantages if we use multiple neural networks:
\begin{itemize}
    \item Compared with modeling the complex global system behavior using a single large-size neural network $\Phi$, a number of neural networks $\Phi_q$, $q=1,\ldots,N$ with much smaller sizes are able to abstract the local system behaviors with the same or even higher modeling precision.
    \item Since $\mathcal{W}_q$, $q=1,\ldots,N$ are independent of each other, the training processes for small-size neural networks $\Phi_q$, $q=1,\ldots,N$  can be conducted in a parallel manner which would be more computationally efficient than training a large scale neural network. 
    \item Even though there are multiple neural networks in the model, only one small-size neural network is activated at each time step $k$. Therefore, the computation effort at each step is only determined by the active small-size neural network. This feature is extremely helpful for executing computationally expensive tasks based on the model such as reachability-based safety verification in which we only need to compute the reachable set of a small-size neural network at each step. 
\end{itemize}

The above benefits of using multiple small-size neural networks enable efficiently verifiable neural-network-based models of dynamical system (\ref{eq:system}). A detailed evaluation will be presented in the evaluation section. 
\end{remark}

After obtaining the collection of neural networks as the set of dynamical descriptions in the hybrid automaton, the transition between two neural networks of dynamical descriptions is defined as follows.

\begin{definition}\label{def_transition}
Transitions between two topologies are automatically generated by the dynamical description of hybrid automaton, such that $\forall p,q=1,2,\ldots,N, \ p\neq q$, $(p,q):x\in\mathcal{X}^{(p)}$, $\Phi_p(x,u)\in\mathcal{X}^{(q)}$.
\end{definition}

\begin{figure}[ht!]
\centering
\vspace{0.5em}
	\includegraphics[width=8cm]{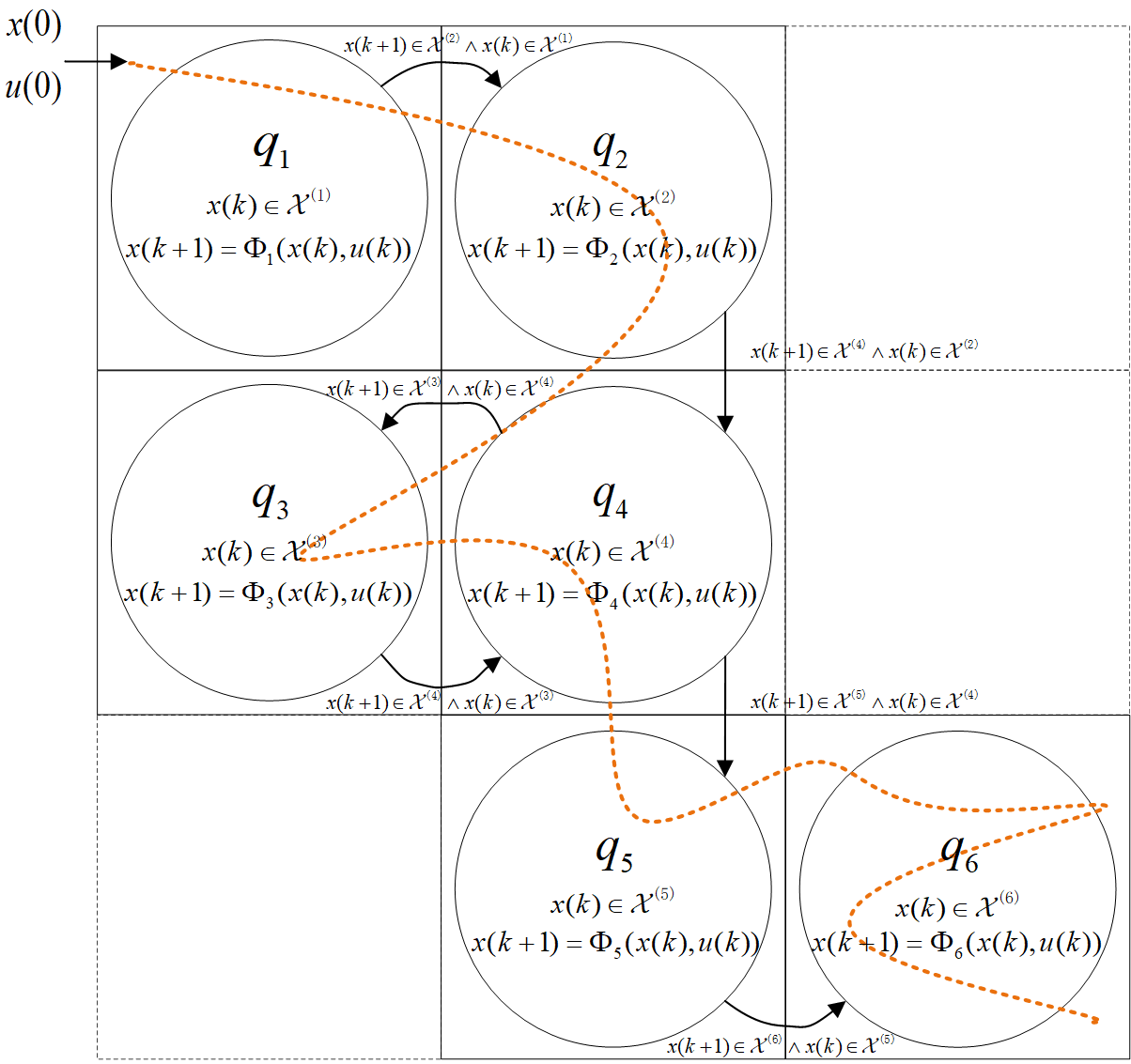}
	\caption{A neural-network-based hybrid automaton model ${\mathcal{H}}$ is constructed with $6$ topologies valid partitions ${\mathscr{X}}=\{\mathcal{X}^{(1)},\mathcal{X}^{(2)},\mathcal{X}^{(3)},\mathcal{X}^{(4)},\mathcal{X}^{(5)},\mathcal{X}^{(6)}\}$ and one state trajectory (in Brown) evolves in $\mathcal{X}$.}
	\label{Fig_Learning automaton} 
\end{figure}

An illustration of neural-network-based hybrid automaton model ${\mathcal{H}}$ is given in Fig. \ref{Fig_Learning automaton}.  After modeling the dynamical system (\ref{eq:system}) in the framework of hybrid automaton model ${\mathcal{H}}$, we can evaluate ${\mathcal{H}}$ using the following Mean Square Error (MSE) performance out of $L$ test traces which is defined as
\begin{align*}\label{ali_MSE}
    \mathrm{MSE}= \frac{1}{\sum\limits_{\ell=1}^{L}M_i}\sum_{\ell=1}^L\left\|\sum_{k=1}^{M_i-1}(\Phi_q(x_i(k),u(k))-x_{i}(k+1))\right\|
\end{align*}
in which $M_i$ denotes the length of $i$th trace. In the evaluation example, this MSE performance will be used for evaluating model precision. 

\subsection{Reachable Set Analysis}

In this subsection, Problem 2, i.e., safety verification, will be addressed in the framework of reachability. The reachable set of hybrid automaton model $\mathcal{H}$ is defined as follows.

\begin{definition} \label{reachable_set}
	Given a neural-network-based hybrid automaton model $\mathcal{H}$  with initial set  $\mathcal{X}_0$ and input set $\mathcal{U}$, the reachable set at a time instant $k$ is $
	\mathcal{X}_{k} \triangleq \{x(k) \mid x(k)~\mathrm{satisfies}~ \mathcal{H}~\mathrm{and}~x(0) \in \mathcal{X}_{0}\} $
	and the reachable set over time interval $[0,k_f]$ is defined by
$\mathcal{X}_{[0,k_f]} = \bigcup\nolimits_{s=0}^{k_f}\mathcal{X}_{s}$.
\end{definition}

Reachable set analysis for a neural-network-based hybrid automaton can be referred for safety verification in \cite{xiang2020reachable}. Based on Definition \ref{reachable_set}, $\mathcal{X}_{k}$ may intersect multiple elements from $\mathscr{X}$, which means there will be a split computation of reachable set for each intersection. This process is called \emph{Split} and is defined by

\begin{definition}\label{def_split}
For a reachable set $\mathcal{X}_{k}$ of $\mathcal{H}$ intersects with $l$ elements of $\mathscr{X}$, given a subspace $\mathcal{V}_{i,k}$ from $\mathcal{X}_{k}$ in which $\mathcal{V}_{i,k}:\mathcal{V}_{i,k}= (\mathcal{X}_{k}\cap\mathcal{X}^{(m)});~ \cup_{i=1}^l\mathcal{V}_{i,k}=\mathcal{X}_{k},~\forall i=1,\cdots,l,~ \exists m=1,\cdots,N$, the process of splitting analysis of the output space $\mathcal{V}_{i,k+1}$ is given by
\begin{align}
   \mathcal{V}_{i,k+1}\triangleq\{\eta_{i,k+1} \mid \eta_{i,k+1}=\Phi_m({\eta_{i,k}}),~ \eta_{i,k}\in\mathcal{V}_{i,k}\}
\end{align}
where the process of obtaining $\mathcal{V}_{i,k+1},~\forall i = 1,2,\cdots,N$ is called \emph{Split}. 
\end{definition}

After $\emph{Split}$, the \emph{Combine} process is needed to obtain a complete reachable set for the next step.
\begin{definition}\label{def_combine}
For $\mathcal{V}_{i,k+1},~i=1,\cdots,l$ the output reachable set $\mathcal{X}_{k+1}$ for a neural-network-based hybrid automaton model $\mathcal{H}$ at time step $k+1$ is given by
$
    \mathcal{X}_{k+1}\triangleq \bigcup_{i=1}^{l}\mathcal{V}_{i,k}
$
by which the \emph{Combine} derives the reachable set at $k+1$ time instance.
\end{definition}

With the \emph{Split and}  \emph{Combine} defined above, the reachable set  of $\mathcal{H}$ can be paralleling analyzed at time instance $k$ if $\mathcal{X}_{k}$ intersects with multiple elements from $\mathscr{X}$. The \emph{Split} and \emph{Combine} is given in Algorithm \ref{alg2}.

\begin{figure}
\vspace{0.5em}
\centering
   \includegraphics[width=6cm]{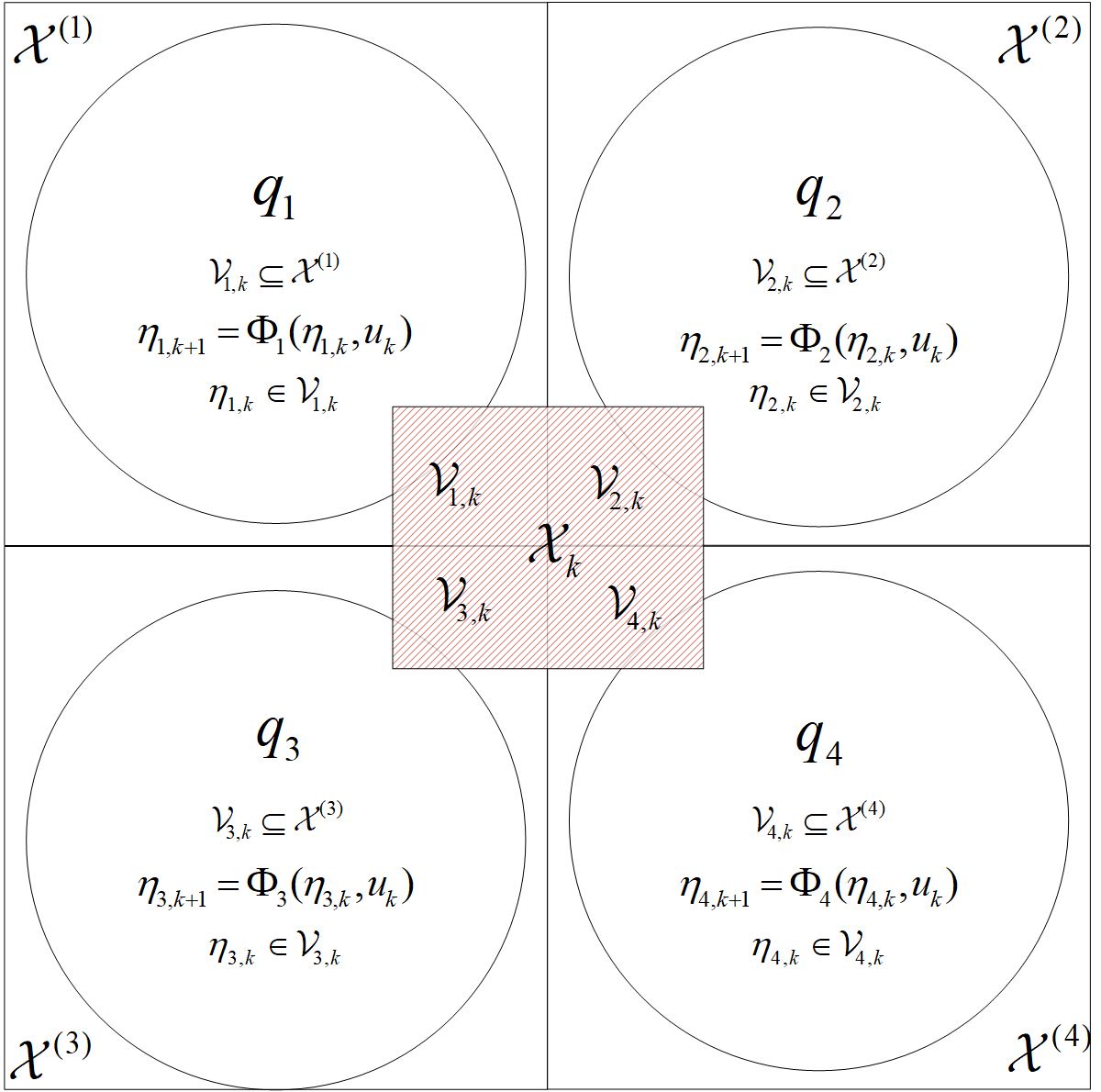}
	\caption{Sketch map for \emph{Split} while $\mathcal{X}_{k}$ intersects with $l=4$ valid partitions}
	\label{Fig_SplitCombine} 
\end{figure}

\begin{algorithm}[ht!] 
\SetAlgoLined
\SetKwInOut{Input}{Input}
\SetKwInOut{Output}{Output}
\Input{Reachable set $\mathcal{X}_{k}$; Neural-network-based hybrid automaton model $\mathcal{H}$.}
\Output{Output Reachable Set $\mathcal{X}_{k+1}$.}
\tcc{Initialization}
$i \gets 1$\\
\tcc{\emph{Split}}
\While{$i\le N$}{

\eIf{$\mathcal{X}^{(i)}\cap \mathcal{X}_{k}\neq\emptyset$}{
$\mathcal{V}_{i,k}\gets \mathcal{X}^{(i)}\cap\mathcal{X}_{k}$\\
Computing $\mathcal{V}_{i,k+1}$ \tcp*{Under Definition \ref{def_split}}
}{$\mathcal{V}_{i,k}\gets \emptyset$}
$i \gets i+1$
}
\tcc{\emph{Combine}}
Computing $\mathcal{X}_{k+1}$ \tcp*{Under Definition \ref{def_combine}}

\Return{$\mathcal{X}_{k+1}$}
\caption{Pseudo Code for \emph{Split and} \emph{Combine}}\label{alg2}
\end{algorithm}

\begin{remark}
Note that if the output reachable set $\mathcal{X}_{k}$ intersects with multiple valid partitions, the process of Split \& Combine may compute the output reachable sets for more times than modeling with a single neural network and the conservatism may increase because of Combine. However, according to \cite{xiang2018output,xiang2020reachable,tran2020nnv}, the computational cost for the set-valued analysis of the neural network is mainly affected by the scale of the neural network model, e.g., the layers and neurons of the neural network model. In our case, due to parallel training of shallow neural network models,  the neural-network-based hybrid automaton model may have less computational cost compared with traditional methods. 
\end{remark}

\section{Evaluation}

In this section, a numerical example of the limit cycle borrowed from \cite{strogatz2001nonlinear} is used for evaluation in the form of
\begin{align} 
    r(k+1)&=(1+\tau)r(k)-\tau r^{3}(k)+\tau u(k)\nonumber\\
    \theta(k+1)&=\theta(k)+\tau\omega\label{ali_limit cycle}\\
    u(k)&=\mu+\delta\zeta(k)\nonumber
\end{align}
where $\omega= 2\pi/3$ and $\tau=0.1$ are the angular velocity and time step width, respectively and the uniform random number $\zeta(k)\sim U(-1,1)$. Namely, the input $u(k)\sim U(\mu-\delta,\mu+\delta)$ ($\mu=0.2$ and $\delta=1.5$) in which $U$ denotes uniform distribution. 

\begin{figure}[ht!]
\centering
	\includegraphics[width=8cm]{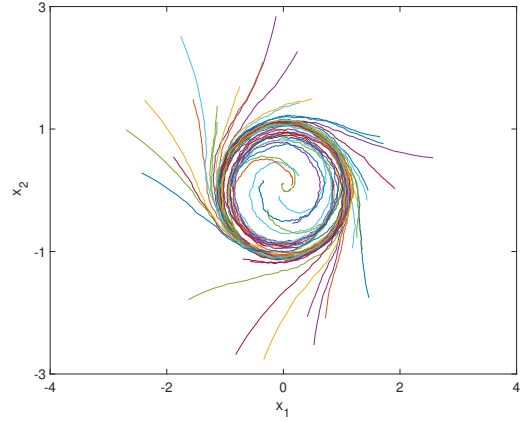}
	\caption{50 trajectories of the limit cycle with random initial condition $r(0)\in [-4,4]$, $\theta(0)\in[-\pi,\pi]$ each of which contains 150 samples and the input $u\sim U(-1.3,1.7)$.}
	\label{Fig_Limitcycle} 
\end{figure}
\begin{figure}[htbp]
\centering
	\subfigure[Single neural network with 200 hidden neurons]{\includegraphics[width=4.1cm]{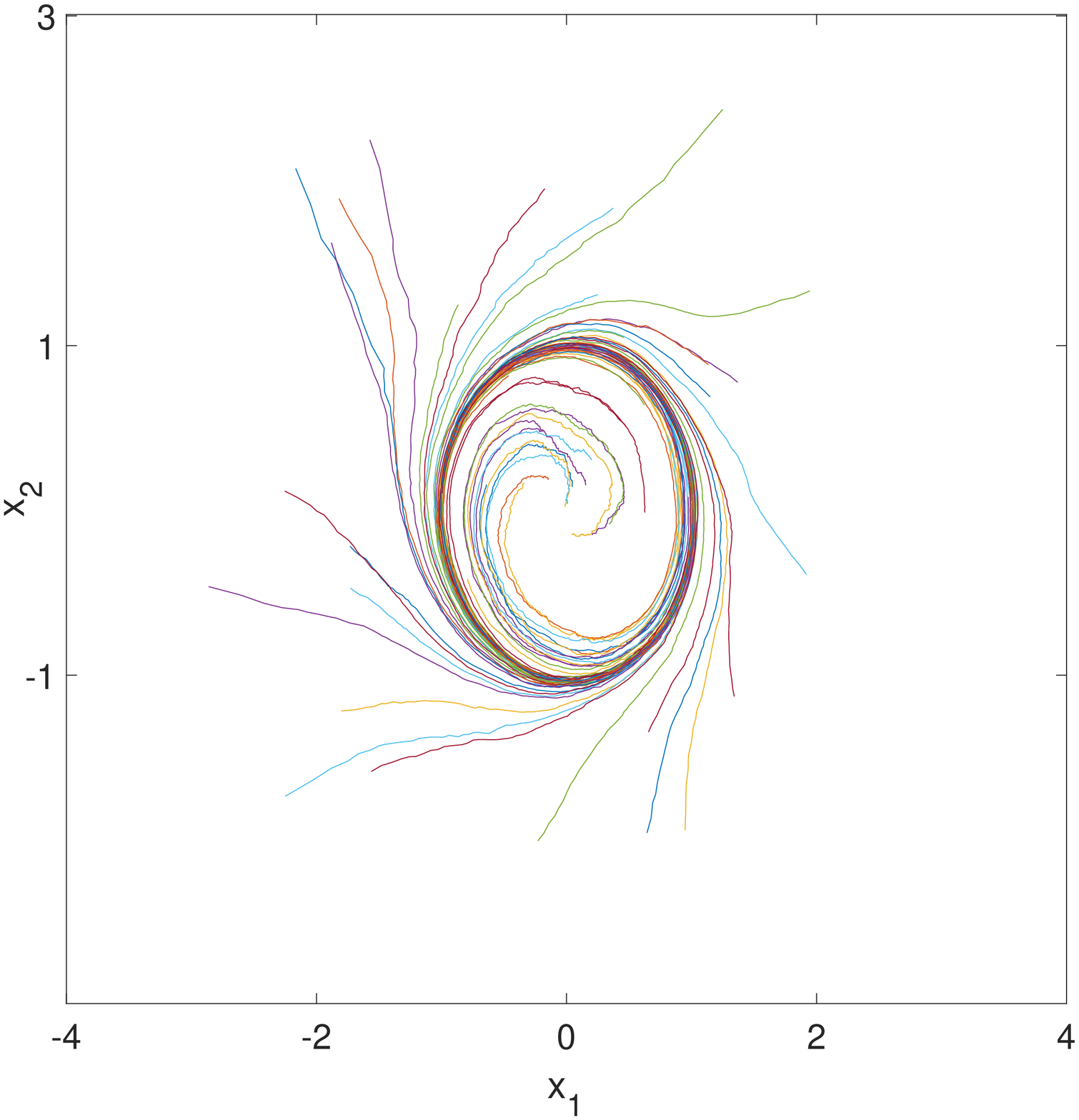}}
	\subfigure[Hybrid automaton $\mathcal{H}$ with 12 neural networks (20 Neurons)]{\includegraphics[width=4.2cm]{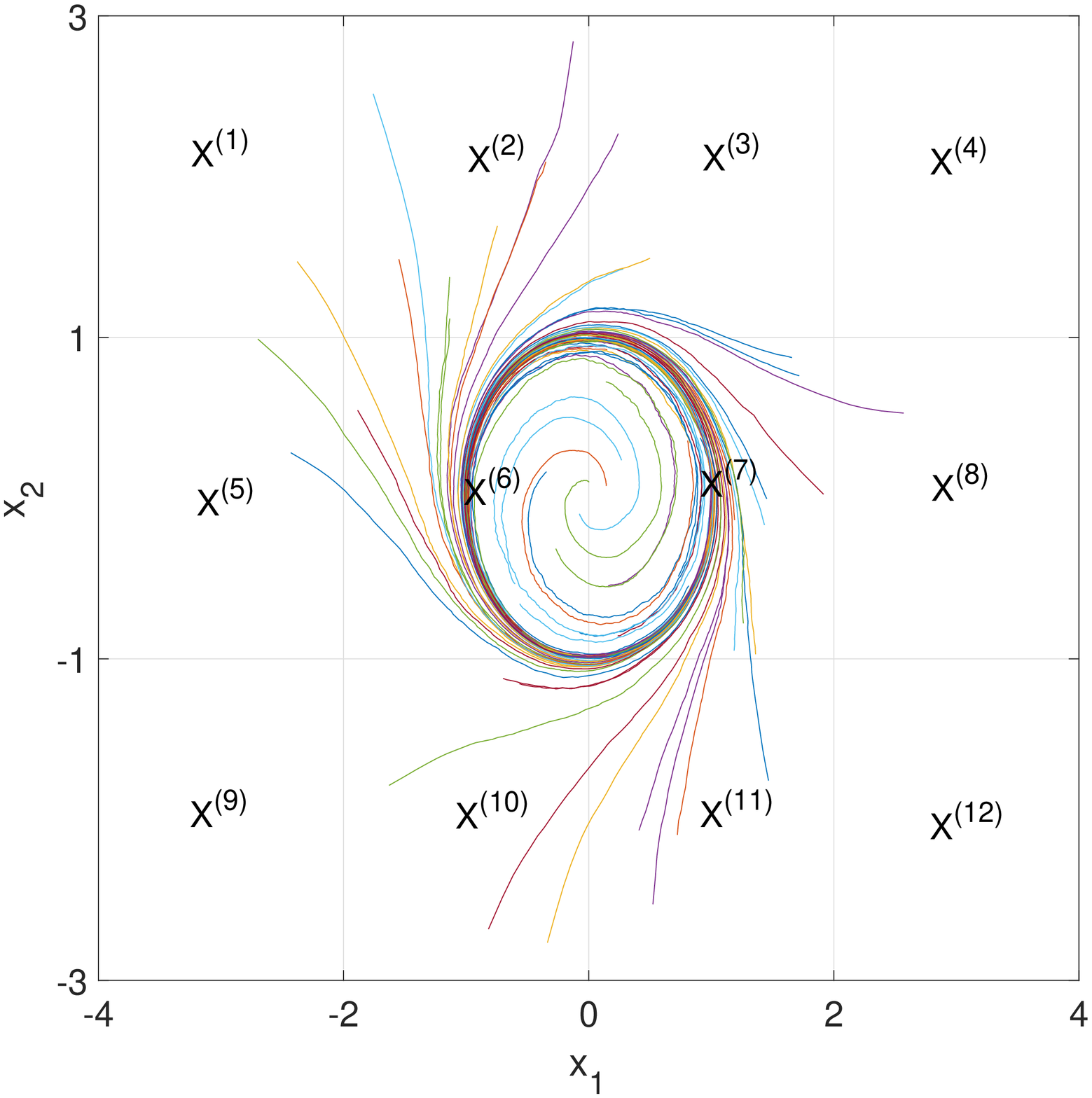}}
	\caption{50 test trajectories of single neural network model (a) and neural-network-based hybrid automaton model (b) with 12  neural networks. }
	\label{Fig_compare} 
\end{figure}
Given dynamical system (\ref{ali_limit cycle}), 50 training traces $\emph{w}=\{w_1,w_2,\cdots,w_{50}\}$ with $h_i=50,~\forall i=1,2,\cdots,50$ are generated with random initial output where $\theta(0)\in[-\pi,\pi]$, $r\in[-4,4]$. 
To derive the neural-network-based hybrid automaton ${\mathcal{H}}$ which aims to model the dynamics of system (\ref{ali_limit cycle}). Firstly, the state region $\mathcal{X}$ defines $x_1\in[-4,4]$ and $x_2\in [-3,3]$. Based on $\mathcal{X}$, valid partitions representing the \emph{Topologies} of ${\mathcal{H}}$ are obtained by desired segments with $M_1=4$, $M_2=3$ for each dimension, totaling 12 topologies. The \emph{Transitions} between topologies are obtained by the relationship between samples and regions.  

\begin{table}[b!]
	\centering
	\caption{Training Time and MSE}\label{pro_1_tab}
	\label{tab1}
	\begin{tabular}{|ccc|}
		\hline 
\textbf{Method}  &\textbf{MSE}& \textbf{Training Time}\\
		\hline\hline
Single Neural Network Model $\Phi$  &$0.0817$  &$4.35$ s  \\
\hline
Hybrid Automaton Model $\mathcal{H}$ &$0.0346$   &$0.32$  s \\
\hline
	\end{tabular}
\end{table} 

Then, by training a set of neural networks with each $\Phi_i$ containing $20$ hidden neurons for each $q_i$, the hybrid automaton model ${\mathcal{H}}$ is obtained. Moreover, for the sake of comparison, a neural network model $\Phi$ with $200$ hidden neurons with similar MSE is trained as well. Fig. \ref{Fig_compare} shows that both single neural network and hybrid automaton models can capture the system's behaviors well. However, in Table \ref{pro_1_tab}, it can be observed that hybrid automaton model ${\mathcal{H}}$ is with lower MSE which implies higher modeling precision.  In addition, the training time can be significantly reduced by training for 12 neural networks parallelly.

\begin{figure}
\centering
	\includegraphics[width=8cm]{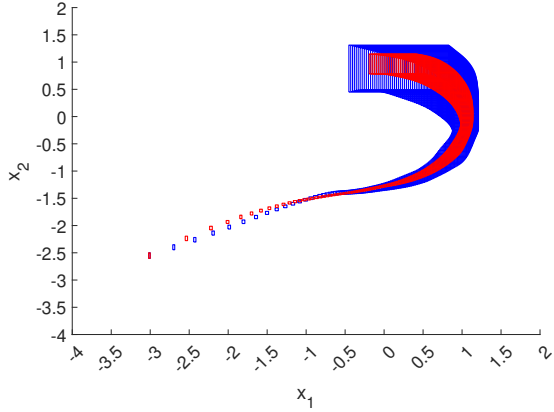}
	\caption{$200$-time-step reachable set computation of $\mathcal{H}$ (red) and a single neural network $\Phi$ (blue) with initial output $x_{1,0}\in[-3.02,-3],x_{2,0}\in[-2.603,-2.5]$ and input $u(k)\in[-1.3,1.7]$}
	\label{Fig_ReachableASM} 
\end{figure}
\begin{figure}
\centering
	\includegraphics[width=8cm]{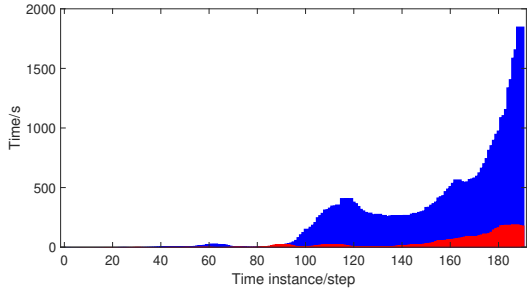}
	\caption{Time for reachable set computation, with a single neural network with 200 hidden neurons (blue), and hybrid automaton $\mathcal{H}$ with 12 neural networks (red).}
	\label{Fig_VerificaitonalTime} 
\end{figure}

Set-valued analysis using NVV in \cite{tran2020nnv} and \emph{Split and}  \emph{Combine} compared with a single neural-network-based dynamical model is given in Fig. \ref{Fig_ReachableASM}, and hybrid automaton model can produce tighter reachable set other than single neural network does. Moreover, the reachable set computation time has been significantly reduced compared with a single neural network model as shown in Fig. \ref{Fig_VerificaitonalTime}. 

In summary, the evaluation results show that the developed neural-network-based hybrid automaton model can enable computationally efficient training and reachability analysis processes with better modeling accuracy and reachability analysis results for complex dynamical systems.


\section{Conclusions}
A data-driven neural-network-based hybrid automaton is developed to model complex dynamical systems. First, sampled data is generated by the dynamical system given random initial conditions and input series. Then, the region of output is divided into valid partitions, based on which the topologies and guards of the proposed model are obtained. Neural networks are trained as the dynamical description for their corresponding topologies. The set-valued reachability of the proposed model is analyzed by the reachable set estimation method and \emph{Split and}  \emph{Combine}. Modeling a numerical example of the limit cycle with a neural-network-based hybrid automaton is given. Compared with the traditional model method with one single neural network, the computational cost can be significantly reduced for computationally expensive tasks such as reachable set computation. 

\bibliographystyle{IEEEtran}
\bibliography{reference.bib}

\begin{thebibliography}{10}
\providecommand{\url}[1]{#1}
\csname url@samestyle\endcsname
\providecommand{\newblock}{\relax}
\providecommand{\bibinfo}[2]{#2}
\providecommand{\BIBentrySTDinterwordspacing}{\spaceskip=0pt\relax}
\providecommand{\BIBentryALTinterwordstretchfactor}{4}
\providecommand{\BIBentryALTinterwordspacing}{\spaceskip=\fontdimen2\font plus
\BIBentryALTinterwordstretchfactor\fontdimen3\font minus
  \fontdimen4\font\relax}
\providecommand{\BIBforeignlanguage}[2]{{%
\expandafter\ifx\csname l@#1\endcsname\relax
\typeout{** WARNING: IEEEtran.bst: No hyphenation pattern has been}%
\typeout{** loaded for the language `#1'. Using the pattern for}%
\typeout{** the default language instead.}%
\else
\language=\csname l@#1\endcsname
\fi
#2}}
\providecommand{\BIBdecl}{\relax}
\BIBdecl

\bibitem{kuptsov2021artificial}
P.~V. Kuptsov, A.~V. Kuptsova, and N.~V. Stankevich, ``Artificial neural
  network as a universal model of nonlinear dynamical systems,'' \emph{arXiv
  preprint arXiv:2104.05402}, 2021.

\bibitem{ghazvini2020experimental}
M.~Ghazvini, H.~Maddah, R.~Peymanfar, M.~H. Ahmadi, and R.~Kumar,
  ``Experimental evaluation and artificial neural network modeling of thermal
  conductivity of water based nanofluid containing magnetic copper
  nanoparticles,'' \emph{Physica A: Statistical Mechanics and its
  Applications}, vol. 551, p. 124127, 2020.

\bibitem{roshni2020neural}
T.~Roshni, M.~K. Jha, and J.~Drisya, ``Neural network modeling for
  groundwater-level forecasting in coastal aquifers,'' \emph{Neural Computing
  and Applications}, pp. 1--18, 2020.

\bibitem{neumann2013neural}
K.~Neumann, A.~Lemme, and J.~J. Steil, ``Neural learning of stable dynamical
  systems based on data-driven {L}yapunov candidates,'' in \emph{2013 IEEE/RSJ
  International Conference on Intelligent Robots and Systems}.\hskip 1em plus
  0.5em minus 0.4em\relax IEEE, 2013, pp. 1216--1222.

\bibitem{he2020learnability}
H.~He, M.~Chen, G.~Xu, Z.~Zhu, and Z.~Zhu, ``Learnability and robustness of
  shallow neural networks learned by a performance-driven {BP} and a variant of
  {PSO}for edge decision-making,'' \emph{Neural Computing and Applications},
  vol.~33, no.~20, pp. 13\,809--13\,830, 2021.

\bibitem{yang2021robust}
Y.~Yang and W.~Xiang, ``Robust optimization framework for training shallow
  neural networks using reachability method,'' in \emph{2021 60th IEEE
  Conference on Decision and Control (CDC)}, 2021, pp. 3857--3862.

\bibitem{fazlyab2019efficient}
M.~Fazlyab, A.~Robey, H.~Hassani, M.~Morari, and G.~Pappas, ``Efficient and
  accurate estimation of {L}ipschitz constants for deep neural networks,''
  \emph{Advances in Neural Information Processing Systems}, vol.~32, 2019.

\bibitem{xiang2018output}
W.~Xiang, H.-D. Tran, and T.~T. Johnson, ``Output reachable set estimation and
  verification for multilayer neural networks,'' \emph{IEEE Transactions on
  Neural Networks and Learning Systems}, vol.~29, no.~11, pp. 5777--5783, 2018.

\bibitem{tran2020nnv}
H.-D. Tran, X.~Yang, D.~Manzanas~Lopez, P.~Musau, L.~V. Nguyen, W.~Xiang,
  S.~Bak, and T.~T. Johnson, ``\textsc{NNV}: the neural network verification
  tool for deep neural networks and learning-enabled cyber-physical systems,''
  in \emph{International Conference on Computer Aided Verification}.\hskip 1em
  plus 0.5em minus 0.4em\relax Springer, 2020, pp. 3--17.

\bibitem{zhang2018efficient}
H.~Zhang, T.-W. Weng, P.-Y. Chen, C.-J. Hsieh, and L.~Daniel, ``Efficient
  neural network robustness certification with general activation functions,''
  \emph{Advances in Neural Information Processing Systems}, vol.~31, 2018.

\bibitem{wang2021beta}
S.~Wang, H.~Zhang, K.~Xu, X.~Lin, S.~Jana, C.-J. Hsieh, and J.~Z. Kolter,
  ``Beta-crown: Efficient bound propagation with per-neuron split constraints
  for neural network robustness verification,'' \emph{Advances in Neural
  Information Processing Systems}, vol.~34, 2021.

\bibitem{beg2017model}
O.~A. Beg, H.~Abbas, T.~T. Johnson, and A.~Davoudi, ``Model validation of
  \textsc{PWM DC-DC} converters,'' \emph{IEEE Transactions on Industrial
  Electronics}, vol.~64, no.~9, pp. 7049--7059, 2017.

\bibitem{xiang2020reachable}
W.~Xiang, H.-D. Tran, X.~Yang, and T.~T. Johnson, ``Reachable set estimation
  for neural network control systems: A simulation-guided approach,''
  \emph{IEEE Transactions on Neural Networks and Learning Systems}, vol.~32,
  no.~5, pp. 1821--1830, 2021.

\bibitem{asarin2007hybridization}
E.~Asarin, T.~Dang, and A.~Girard, ``Hybridization methods for the analysis of
  nonlinear systems,'' \emph{Acta Informatica}, vol.~43, no.~7, pp. 451--476,
  2007.

\bibitem{bak2016scalable}
S.~Bak, S.~Bogomolov, T.~A. Henzinger, T.~T. Johnson, and P.~Prakash,
  ``Scalable static hybridization methods for analysis of nonlinear systems,''
  in \emph{Proceedings of the 19th International Conference on Hybrid Systems:
  Computation and Control}, 2016, pp. 155--164.

\bibitem{strogatz2001nonlinear}
S.~H. Strogatz, \emph{Nonlinear Dynamics and Chaos: with Applications to
  Physics, Biology, Chemistry, and Engineering}.\hskip 1em plus 0.5em minus
  0.4em\relax CRC press, 2018.

\end{thebibliography}

\end{document}